\documentclass[11pt,onecolumn,amssymb,nofootinbib]{revtex4}
\usepackage{amsmath, amsthm, amscd, amssymb}
\usepackage{graphicx, braket}
\usepackage{bm}
\usepackage{bbm}
\usepackage{epstopdf}

\begin{document}

\title{\bf Hidden Sector Dark Matter Realized as a Twin of the Visible Universe With Zero Higgs Vacuum Expectation } \bigskip

\author{Stephen L. Adler}
\email{adler@ias.edu} \affiliation{Institute for Advanced Study,
1 Einstein Drive, Princeton, NJ 08540, USA.}

\begin{abstract}
We propose that the universe contains two identical sets of particles and gauge interactions, coupling only through gravitation, which differ by their Higgs potentials.   We postulate that because of underlying symmetries, the two sectors when uncoupled  have Higgs potentials that lie  at the boundary between phases with nonzero and zero Higgs vacuum expectation.  Turning on the coupling between the two sectors can break the degeneracy, pushing the Higgs potential in one sector  into the domain of nonzero Higgs expectation (giving the visible sector), and pushing the Higgs potential in the other sector  into the domain of zero Higgs expectation (giving the dark sector).  The least massive baryon in the dark sector will then be a candidate self-interacting dark matter particle.

\end{abstract}
\leftline{Essay written for the Gravity Research Foundation 2024 Awards for Essays on Gravitation}
\maketitle
\section{Dark matter}
There is now abundant evidence, both from galactic rotation curves and from studies of the cosmic microwave background, that ordinary baryonic matter makes up only about one sixth of the matter content of the universe, with the rest consisting of a mysterious form of ``dark matter''.  To match observations, the dark matter should have at a minimum the following three properties: (i) It should gravitate like normal matter, but to be ``dark'' it should have zero or very weak couplings to ordinary matter, (ii) In the early universe, it should be formed at an order of magnitude similar rate to ordinary matter, (iii) Its  aggregation properties should be quite different from those of ordinary matter.

So far, much attention has been devoted to the possibility that dark matter has nonzero but very weak couplings to ordinary matter, with condition (ii) obeyed by appropriate choices of dark matter mass and couplings to visible matter (as in the ``WIMP miracle''.)
Direct searches for dark matter are ongoing, but so far searches for weakly interacting massive particles  \cite{wimp1}, \cite{wimp2} have  produced only  upper bounds, while searches for axion dark matter \cite{axion} are mostly just starting.    Whether current or future experiments continue to give only bounds,  or whether dark matter direct detection lies in the near future, will be seen over the next few years.

The lack  of direct detection of dark matter up to the present has prompted suggestions that perhaps all or most of the dark matter lies in a ``hidden sector'' of the universe, with only couplings to gravitation, which is all that is needed to explain the astrophysical and cosmological evidence for dark matter. A natural way to satisfy condition (ii) in this context is to have a dark sector that differs by a discrete reflection symmetry from the visible sector.   There have been extensive investigations of the idea that dark matter is ``mirror matter'', with spatial parity properties opposite to those of ordinary matter; for example, left chiral gauge interactions of ordinary matter would be replaced by right chiral ones  of mirror dark matter \cite{wiki}, \cite{foot}, \cite{footrev}.  To restore overall parity invariance of the universe one requires the two sectors to have Lagrangians that are  reflection images of one another. This necessitates a  mechanism, specifically  \cite{footrev}  asymmetric initial conditions, to explain the observed differences between the visible and dark sectors of the universe, as required by condition (iii).

An alternative suggestion for creating a hidden sector differing by a reflection symmetry, which we first made in the context of our exposition \cite{td} of ``trace dynamics'' as a pre-quantum theory, and elaborated on afterwards in a Gravitation Essay \cite{essay}, is to have a dark sector in which the imaginary unit, that is taken by convention as $+i$ in the visible sector, has the opposite sign $-i$ in the dark sector.  Again,  one needs a  mechanism to explain the observed differences between the visible and dark sectors of the universe.

Another extensively studied idea for hidden sector dark matter focuses on ``dark photons'', that is, additional $U(1)$ gauge interactions \cite{lagouri}, with coupling requirements imposed by conditions (i)-(iii).

In this Essay we propose an alternative doubling mechanism for generating hidden sector dark matter, based on the observation that the origin of the ``wrong sign'' negative quadratic term in the standard model Higgs potential is still mysterious, suggesting that it may be paired with another sector of particles where this quadratic term has the normal positive sign. As argued in the next section, this can give a natural mechanism for satisfying conditions (i)-(iii), while leading to self-interacting dark matter studied by Spergel and Steinhardt \cite{spergel}, which solves a number of problems associated with the original WIMP idea.\footnote{For succinct summaries of the problems with the standard WIMP scenario, see Sec. 2 of \cite{nick} and Secs. II\,A--II\,D of \cite{tulin}.  Recent observations finding more faint dwarf galaxies, reviewed in \cite{simon}, can alleviate the ``missing satellite'' problem, and so-called baryon feedback processes, reviewed in Sec. II\,E of \cite{tulin}, can alleviate the ``core-cusp''problem.  Nonetheless, there is currently strong interest in the idea of self-interacting dark matter, as exemplified by the reviews \cite{nick}, \cite{tulin}, \cite{sarkar}, and \cite{adhikari}.}

\section{A Higgs mechanism for two copies of quantum field theory to give the visible universe and the dark universe}

The standard model potential for the Higgs field $\phi$ can be conveniently written in the form
\begin{equation}\label{pot}
V(\phi)=\mu^2 \phi^*\phi +  \lambda (\phi^* \phi)^2~~~.
\end{equation}
When $\mu^2$ and $\lambda$ are both positive, the classical minimum lies at $|\phi|=0$,  while when $\mu^2<0$ and $\lambda$ is positive, the classical minimum lies at the nonzero value $|\phi|=\big(-\mu^2/(2 \lambda)\big)^{1/2}$.  When radiative corrections are included, $\lambda$ becomes itself a function of $|\phi|$, and even when positive at small $|\phi|$ can run to metastable or  unstable negative values at large $|\phi|$.  Recent determinations of  the Higgs boson mass  measured by Atlas \cite{atlas} as $125.22 \pm 0.14$ GeV, together with the top quark mass as measured by the Tevatron \cite{tevatron} as $172.76 \pm 0.3$ GeV, and by CMS \cite{cms} as $171.77 \pm 0.38$ GeV, puts the standard model vacuum on the cusp of metastability \cite{alekhin}, \cite{degrassi}, \cite{espinosa}.  This may be an accident, or an  indication of missing new physics \cite{ellis}  beyond the standard model that stabilizes  the vacuum. But taken at face value, the measured Higgs and top quark masses give  an indication that there is something very special about the parameters characterizing the standard model vacuum, and it is possible that the vacuum lies at a finely tuned boundary between stability and instability \cite{arvanblog}.

Partly motivated by this, we postulate that in a two-copy universe, before coupling of the copies through  gravitation, a selection rule in an underlying theory places both copies at the $\mu^2=0$ boundary between the Higgs phases with nonzero and zero vacuum expectation.
It is then plausible for coupling of the two  copies through gravitation to move the copies in opposite directions (analogous to level repulsion in perturbation theory), with one copy moving into the domain of nonzero Higgs vacuum expectation ($\mu^2 <0$), and the second copy moving into the domain of zero Higgs vacuum expectation ($\mu^2>0$).  The copy with a nonzero vacuum expectation of the Higgs field will then give the visible universe, and the second copy, in which the vacuum expectation value of the Higgs field vanishes, will give a candidate for the dark universe.\footnote{Note that the second copy still has a full electroweak gauge group, so our proposal differs in this respect from the universe without weak interactions discussed in \cite{weakless}.}

What would this dark universe look like?  Studies of standard model-like theories with either vanishing Higgs vacuum expectation, or no Higgs sector altogether, have been given in a blog by Strassler \cite{strass}, in a detailed article of Quigg and Shrock \cite{quigg}, and in on-line Physics Stack Exchange  discussions \cite{stack}.  The dark sector Higgs particle would remain massive, but its rapid decay would make it irrelevant for astrophysical dark matter. So to assess the dark matter implications of a sector with vanishing Higgs expectation, it suffices to study models with no Higgs sector, as in \cite{quigg}.

Even with vanishing Higgs vacuum expectation, dynamical symmetry breaking through condensate formation \cite{quigg},  \cite{wilczek} in a dark sector analog of QCD will give masses to baryons.
The lightest nucleon analog, which may be charged or neutral \cite{quigg}, would still have a mass of order a GeV,\footnote{This assumes the chromodynamics running couplings in the hidden and visible sectors are the same.} and so easily satisfies the Tremaine-Gunn \cite{tg} lower bound of order 100 eV on fermionic dark matter masses, and can act as a kind of WIMP.    If the lightest baryon is neutral, no dark matter ``atoms'' will form, and so dark matter aggregation properties will be very different from those of visible matter.  If the lightest baryon is charged, the form of dark matter atoms depends on the lepton masses.  Since the source of standard model lepton masses is their coupling to the Higgs field vacuum expectation, leptons in a Higgs expectation-less world remain massless at tree level. In the absence of other sources of lepton mass, dark atoms again would not form.  In certain models of electroweak dynamical symmetry breaking discussed in Sec. V of \cite{quigg}  (see  also \cite{piai}, \cite{mini}), the leptons can attain dynamical masses much smaller than an  eV.\footnote{This result depends on the ultraviolet completion used in \cite{quigg}; much larger masses can be obtained with alternative ultraviolet completions.}  Dark atoms would than have an exponentially larger Bohr radius than visible sector atoms, and so again would have very different aggregation behavior from visible matter.

The lightest baryon in a Higgs expectation-less world would have self-interactions arising both from its QCD analog interactions, and from its coupling to dark sector electroweak gauge bosons, of estimated \cite{quigg} range $(30 \,{\rm MeV})^{-1}$.  The self-interaction cross section of the lightest baryon  will be similar to that of ordinary protons and neutrons, and can satisfy the requirements for the cross section to mass ratio $\sigma/m$ of self-interacting dark matter quoted in  \cite{spergel}, \cite{nick}, \cite{tulin}, \cite{sarkar},  \cite{adhikari}. More recently, Girmohanta and Shrock  \cite{GS1}, \cite{GS2} have studied a model with dark fermions of mass of order 5 GeV, interacting via a mediator of mass 5 MeV.  (See also related work in \cite{yang}.)  They find that this can accommodate fits to astrophysical data, such as that of Kaplinghat, Tulin, and Wu \cite{kaplinghat}, whose ``results prefer a mildly velocity-dependent cross section, from $\sigma/m \simeq 2 \,{\rm cm}^2/{\rm gm}$  on galaxy scales to $\sigma/m \simeq 0.1 \,{\rm cm}^2/{\rm gm}$  on cluster scales''.\footnote{A useful conversion of units is  $1 \,{\rm cm}^2/{\rm gm} \simeq 2\, {\rm barn}/{\rm GeV}$.}     These cross sections are again similar to normal hadronic cross sections, and may well be attained in a Higgs expectation-less dark sector.

\section{Summary}
We suggest that dark matter arises from a hidden sector double of the standard model, with $\mu^2 >0$ in its Higgs potential,  so that  the Higgs field vacuum expectation vanishes, in contrast to the visible sector which has $\mu^2<0$.    Further analysis of this proposal requires only known physics of the standard model QCD and electroweak forces, and so should permit detailed comparisons with observed astrophysical dark matter properties.  We remark finally that while from a minimal, ``bottom up'' perspective, doubling the particle and field content of the universe may seem like a stretch, there have been proposals that this is precisely what happens when fundamental theory is constructed from a ``top down'' viewpoint. Specifically,
there have been suggestions that in the $E_8 \times E_8$ heterotic string,  ``the $E_8 \leftrightarrow E_8$ symmetry may persist even in the dimensionally-reduced theory.... then there would be two forms of matter: ordinary matter, whose interactions would be described by $E_8$, and shadow matter; whose interactions would be described by $E_8$; these two kinds of matter would only interact through gravitational strength interactions \cite{kolb}.''  And in a similar vein, ``the $SU(3) \times SU(2) \times U(1)$ gauge group of the standard model can fit quite nicely within one of the $E8$ gauge groups.   The matter under the other $E8$ would not interact except through gravity, and might provide an answer to the Dark Matter problem in astrophysics \cite {pierre}.''      And foreseeing the QCD-like aspect of our proposal, ``Thus, if the string scale is comparable to the Planck scale, the existence of light fields carrying nontrivial quantum numbers of the second $E_8$ could only be detected by gravitational strength interactions.  These fields comprise the {\it hidden sector}. A hidden sector could actually be useful.  Assuming that the hidden sector has a mass gap, perhaps due to confinement, one intriguing possibility is that hidden-sector particles comprise a component of the dark matter \cite{becker}.''

\section{Acknowledgement}
  This article  benefited from the hospitality of the Aspen Center for Physics, which is supported by National Science Foundation Grant PHY-1607611.  I wish to thank Robert Shrock for helpful comments on my first draft, and the suggestion of relevant references.

\end{document}